\documentclass[
preprint,
amsmath,amssymb,
floatfix,
]{revtex4-2}

\usepackage{graphicx}
\usepackage{dcolumn}
\usepackage{bm}
\usepackage[utf8]{inputenc}
\usepackage{natbib}
\usepackage{float}
\usepackage{amsmath}
\usepackage{caption}
\usepackage{fancyhdr}
\usepackage[dvipsnames]{xcolor}
\usepackage{booktabs}
\usepackage{easyReview}
\usepackage{soul}
\usepackage{mathtools}
\usepackage{makecell}

\begin{document}

\title{Localising two sub-diffraction emitters in 3D using quantum correlation microscopy}%

\author{Shuo Li\textsuperscript{1}, Wenchao Li\textsuperscript{2},  Qiang Sun\textsuperscript{1}, Bill Moran\textsuperscript{2,3}, Timothy C. Brown\textsuperscript{4}, Brant C. Gibson\textsuperscript{1}, Andrew D. Greentree \textsuperscript{1,*}}

\address{1 Australian Research Council Centre of Excellence For Nanoscale Biophotonics, RMIT University, Melbourne 3001, Australia\\
2 School of Science, RMIT University, Melbourne, VIC 3001, Australia\\
3 Department of Electrical and
Electronic Engineering, University of Melbourne, VIC 3010, Australia\\
4 School of Mathematics, Monash University, Melbourne, VIC 3800, Australia}

\email{shuo.li3@rmit.edu.au; andrew.greentree@rmit.edu.au}


\begin{abstract}
The localisation of fluorophores is an important aspect of the determination of the biological function of cellular systems.  Quantum correlation microscopy (QCM) is a promising technique for providing diffraction unlimited emitter localisation that can be used with either confocal or widefield modalities.  However, so far, QCM has not been applied to three dimensional localisation problems. Here we show that quantum correlation microscopy provides diffraction-unlimited three-dimensional localisation for two emitters within a single diffraction-limited spot. By introducing a two-stage maximum likelihood estimator, our modelling shows that localisation precision scales as $1/\sqrt{t}$ where $t$ is the total detection time. Diffraction unlimited localisation is achieved using both intensity and photon correlation from Hanbury Brown and Twiss measurements at as few as four measurement locations. 
\end{abstract}


\maketitle

\section{Introduction}
\label{sec:intro}
The quest for super-resolution imaging and localisation of fluorophores is one of the most important challenges for optical microscopy \cite{betzig2006science, hell2015review, koenderink2022nano, heintzmann2017chemreview,tenne2019np},  a convenient imaging modality for both functional and live-cell imaging.  However, the length scale for visible light restricts imaging to of order 500~nm because of  the diffraction limit.  This length scale is large compared with many biological regions of interest, and this limitation has spurred research into techniques that go beyond the diffraction limit \cite{monticone14prl, bartels2022ic,nehme2018optica}, and are ideally diffraction \emph{unlimited} \cite{klauss2017springer, pascucci2019nc}. 

As well as improving resolution, capturing the three-dimensional structure of the cellular and sub-cellular environment is also vital to understanding its function.  Several techniques exist for three-dimensional imaging, some of which incorporate super-resolution techniques in whole or in part.  Examples of three-dimensional imaging techniques are z-stack \cite{Aga1984,beaulieu2020natmethods,abrahamsson2013natmethods}, light sheet microscopy \cite{stelzer2015natmeth,swoger2014coldspring,stelzer2021natreview} and total internal reflection fluorescence microscopy \cite{szalai2021nc,axelrod2001traffic,boulanger2014pans}.  Both light sheet and total internal reflection fluorescence microscopy require complex optical configurations/components. 

Our goal has been to study the role that quantum effects play on resolution.  In particular, we have been studying quantum correlation microscopy (QCM) \cite{worboys20pra,HSH1995,israel2017nc}.  This approach uses both intensity and photon correlation from Hanbury Brown and Twiss measurements (HBT), to improve localisation and/or imaging, and has  been shown to be useful in both widefield \cite{schwartz2013nl, schwartz2012pra} and confocal \cite{monticone14prl, simon2010oe} imaging modalities. Our approach has more in common with z-stack in that we consider a confocal arrangement using discrete focal points in three-dimensional space.

We first develop the analytics of three-dimensional localisation using quantum correlation microscopy, based on which we develop a maximum likelihood estimator (MLE) using an approximated likelihood function. We confirm our results using Monte-Carlo (MC) simulations for the diffraction unlimited two-emitter localisation process.  
Our methodology allows the efficient usage of information from each photon to minimize the detection time and total intensity, and thereby to mitigate possible phototoxicity observed in super-resolution approaches \cite{IWW+2017}. Our method is particularly suited for high photostability single-photon emitters such as nitrogen-vacancy (NV) centres in diamond as biomarkers \cite{SCL+2014}.

This paper is organised as follows.  First we describe the theory of localisation of two emitters in three dimensions, using both intensity and HBT information. We then develop a MLE estimator to realize the 3D localisation algorithm 
Using the MLE, we run MC simulations for the  3D localisation of  two randomly generated emitters, including different detection configurations. Finally we exhibit  the scaling law of resolution,  demonstrating the diffraction-unlimited potential of the approach.

\section{System}
\label{sec:sys}
\begin{figure}[tb]
  \centering
  \includegraphics[width=0.9\textwidth]{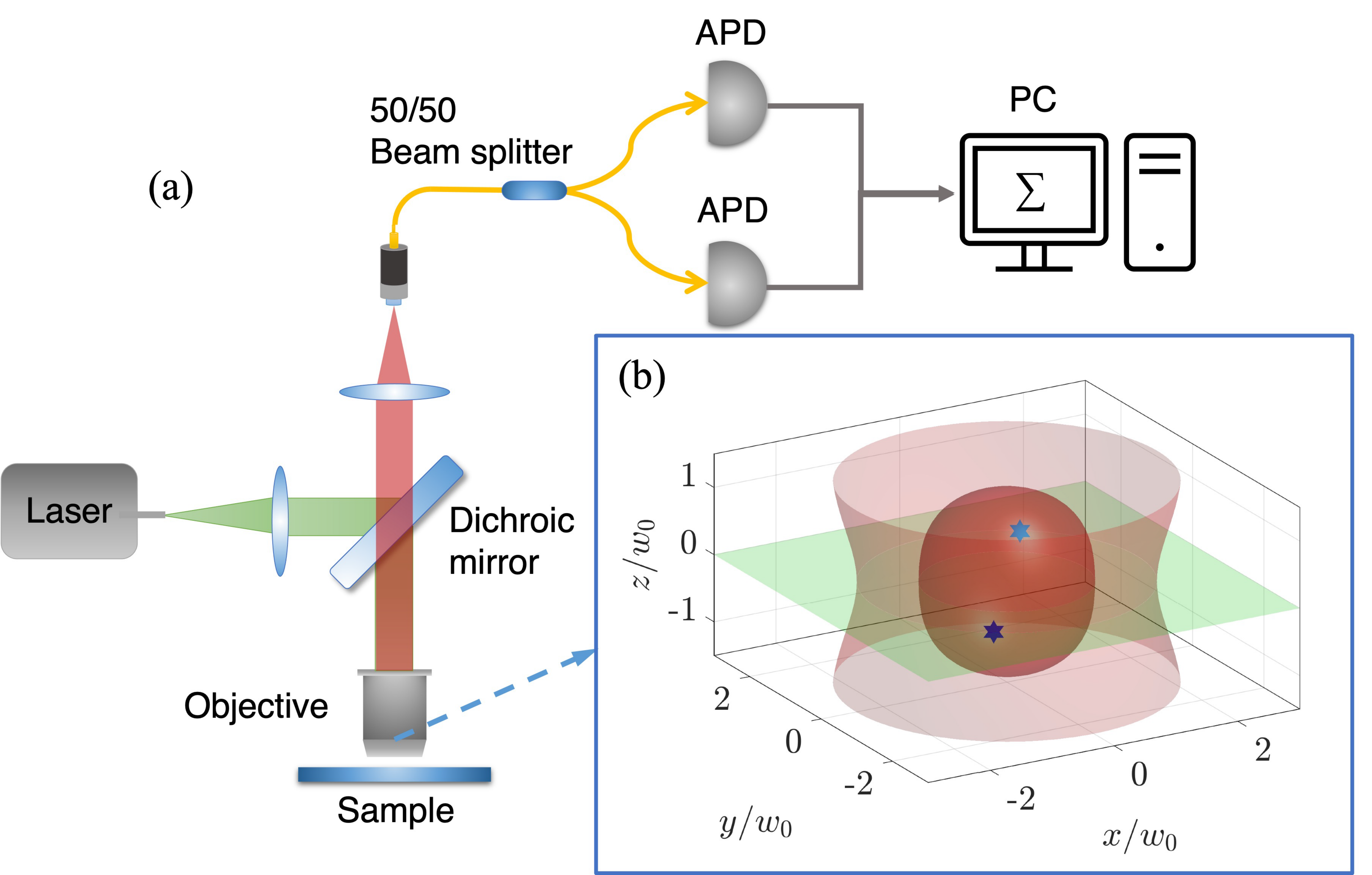}
\caption{(a) Schematic of HBT system of HBT localisation system. (b) Two emitters (stars) within one 3D Gaussian PSF, which corresponds to the photon collection probability.  Excitation is assumed to be uniform. The green plane indicates the objective focal plane, and the red surfaces are the equal intensity surfaces: the outside one is the intensity $I=1/e^2 \sum_i^n{P_{0,i}}$ to show the 3D Gaussian beam waist, the inside one is the intensity $I=0.5 \sum_i^n{P_{0,i}}$ to show the 3D focal spot. Through the confocal microscopy photons from the two emitters are directed via a 50/50 beamsplitter into two APDs, and both intensity and HBT signal are obtained.}
\label{fig:sys}
\end{figure}

We consider two single photon emitters of unequal brightness, positioned in three dimensions within one diffraction limit. Fig.\ref{fig:sys} shows the schematic of 3D HBT measurement.   The emitter positions are  at $\mathbf{x}_1 = (x_1,y_1,z_1)\in\mathbb{R}^3$, $\mathbf{x}_2= (x_2,y_2,z_2)\in\mathbb{R}^3$. For simplicity, we assume that the emitters are excited via an external light source so that the two emitters are equally illuminated.  Collection is performed by a single confocal microscope objective that is capable of being translated in the $x,\, y$, and $z$ directions, which leads to translation of the focal point. We define the focal points as $\mathbf{\xi_j}\in\mathbb{R}^3$, where $j$ ranges of over all of the focal points used for a particular localisation scheme.  This configuration means that the probability of exciting each emitter is equal, though the probability of collecting photons from the emitters varies depending on the precise location of each emitter within the microscope point spread function.

Photons from the two emitters are directed via a 50/50 beamsplitter to two single photon counting modules, e.g. avalanche photodiodes (APD), as shown in Fig.~\ref{fig:sys}. The measurements  from the APDs are read out both independently, to obtain intensity, and in coincidence, to obtain the HBT signal. Assuming each single-photon-emitter emits only one photon during its emission lifetime, correlation events can only occur when each emitter fires one photon and these  are received by the two APDs simultaneously. Both  the intensity and second-order correlation function of the two-emitter system at zero time delay $g^2(\tau=0)$ are used for the two-emitter localisation. 

We define the intrinsic brightness of emitter $i$ to be $P_{0,i}$; this is the maximum probability of detecting a photon from the emitter when it is located at the centre of the detection point spread function (PSF), i.e. when $\mathbf{x}_i = \mathbf{\xi}_j$ for one of the $\mathbf{\xi}_j$. The probability of detecting a photon from emitter $i$ is defined by the microscope point spread function, which we treat as Gaussian
\begin{align}
    P_{i} = \frac{2P_{0,i}}{\pi w(z)^2} \exp\left(-2 \frac{r_i^2}{w(z)^2}\right)
\end{align}
where $r_i = \sqrt{x_i^2 + y_i^2}$ is the transversal distance from emitter $i$ to the detection optical axis, $z$ is the distance propagated from the plane where the wavefront is assumed flat. Note that this definition of the detection probability also takes into account any additional loss processes in the system.  

The beam waist $w(z)$ is the radius of the laser beam at which the irradiance is $1/e^2$ (13.5\%) of the centre of the beam at $z$:
\begin{align}
    w(z) = w_0 \sqrt{1 + \frac{z}{z_R}}, 
\label{eq:wz}
\end{align}
where $w_0$ is the Gaussian beam waist in the focal plane and the Rayleigh range, $z_R$ is the value of $z$ where the cross-sectional area of the beam is doubled:
\begin{align}
    z_R = \frac{\pi w_0^2}{\lambda}.
\end{align}

The coincidence rate is expressed as a function $g^{(2)}(\tau)$, where  $\tau$ is the time delay between photon detections at each of the APDs.  For our purposes we are solely interested in the second order correlation at zero time delay $\tau=0$ (i.e. simultaneous detection at each APD) as this provides the most quantum information about the relative brightness of the emitters:
\begin{align}
    g^{(2)}(\tau=0) = \frac{2 P_1 P_2}{(P_1 + P_2)^2} = \frac{2 \alpha}{(1 + \alpha)^2}
\end{align}
where $\alpha = P_2/P_1$ is the brightness ratio of two emitters ($0<\alpha<1$). In what follows  we use $g^{(2)}$ as  simplified notation for  $g^{(2)}(\tau=0)$.

Each measurement includes an intensity and $g^{(2)}$. These are both functions of emitter coordinates and relative brightness.  Since we are inferring seven independent variables $x_1,y_1,z_1,x_2,y_2,z_2,\alpha$ in total, at least eight equations, that is,  four measurement locations are required to solve the problem, i.e. the minimum number of $\mathbf{\xi_j}$ is $j=1...4$. To reduce the possibility that the 3D searching descends into a local minimum, we introduce a scanning function to localise the maximum intensity area, with which we run an intensity scan in 3D and select the top two maximum intensity locations as initial guesses for the optimizer.

\section{Maximum likelihood localisation}
\label{sec:MLE}

We model the detected photon counts from emitter $i$ via a Poisson distribution
\begin{align}
    c_i &\sim \operatorname{Pois}\left(P_i t\right)
\end{align}
where \textit{t} is the detection time i.e. the photon acquisition time, and $\operatorname{Pois}\left(P_i t\right)$ is the Poisson distribution with rate $P_i t$.

The total normalised intensity from the two emitters can then be expressed as:
\begin{align}
    I = \frac{c_1+c_2}{P_{0,1}t+P_{0,2}t}
\end{align}

The 2nd order correlation is:
\begin{align}
    g^{(2)} =\frac{2c_{12}}{c_{11} + c_{22} + 2 c_{12}}
\end{align}
where $c_{12}$ is the detected photon counts from both emitters simultaneously (coincidence), $c_{11}$ and $c_{22}$ are the uncorrelated events generated by the same emitter 1 or 2,
\begin{equation}
\begin{aligned}
c_{11} &\sim\operatorname{Pois}\left(P_1^2 t\right) \\
c_{22} &\sim\operatorname{Pois}\left(P_2^2 t\right) \\
c_{12} &\sim\operatorname{Pois}\left(P_1 P_2 t\right)
\end{aligned}
\end{equation}

In this paper, the maximum likelihood estimator (MLE) is used to estimate the positions of emitters $\mathbf{x}_1=[x_1,y_1,z_1]\in\mathbb{R}^3$, $\mathbf{x}_2=[x_2,y_2,z_2]\in\mathbb{R}^3$ via measurements $I$ and $g^{(2)}$. Specifically, let $P_{0,1}=1$, $P_{0,2}=\alpha\in(0,1)$, $f_{I_j}=f_{I}(\mathbf{x}_1,\mathbf{x}_2,\bm{\xi}_j,\alpha)$ and $f_{g^{(2)}_j}=f_{g^{(2)}}(\mathbf{x}_1,\mathbf{x}_2,\bm{\xi}_j,\alpha)$ be the probability density functions (PDF) of $I_j$ and $g_j^{(2)}$, respectively,  given the locations of emitters $\mathbf{x}_1$ and $\mathbf{x}_2$, and the focusing positions, i.e. the maximum of PSFs of  the $j$-th ($j=1,\, \cdots,\, m$) detector locations, $\bm{\xi}_j\in\mathbb{R}^3$. For simplicity of notation, $g^{(2)}$ and $g_j^{(2)}$ is written as $g$ and $g_j$, respectively, when they are used as subscripts, e.g. $f_{g^{(2)}}=f_{g}$ and $f_{g_j^{(2)}}=f_{g_j}$. 

Then,  the joint log-likelihood function given the  measurements is
\begin{align}
\ell(\mathbf{x}_1,\mathbf{x}_2,\alpha|\bm{\xi}_j,I_{1:m},g^{(2)}_{1:m}) =& \log\left\{\prod_{j=1}^mf_{I_j}f_{g_j}\right\}\notag\\
    =&\sum_{j=1}^m\Big(\log f_{I_j}(\mathbf{x}_1,\mathbf{x}_2|\bm{\xi}_j,I_j)+\log f_{g_j}(\mathbf{x}_1,\mathbf{x}_2|\bm{\xi}_j,g^{(2)}_j)\Big),
\end{align}
and, as a result, the MLE is
\begin{align}
\ell(\hat{\mathbf{x}}_1,\hat{\mathbf{x}}_2,\hat{\alpha})=\arg\max\ell(\mathbf{x}_1,\mathbf{x}_2,\alpha|\bm{\xi}_j,I_{1:m},g^{(2)}_{1:m}).  
\label{fullmle}
\end{align}

The  PDFs of $I$ and $g^{(2)}$ are not available in closed form,  so the analytic solution to Eq.\eqref{fullmle} is intractable. However these PDFs  approach a Gaussian distribution with increasing $t$ (see  Section~\ref{sec:fim} for more detail). As a result, the likelihood function is approximated by
\begin{align}
\ell(\mathbf{x}_1,\mathbf{x}_2,\alpha|\bm{\xi}_j,I_{1:m},g^{(2)}_{1:m}) \approx \sum_{j=1}^{m}&\left\{ \frac{\left(\mu_{I_j}(\mathbf{x}_1,\mathbf{x}_2,\alpha) - I_j \right)^2}{\sigma^2_{I_j}(\mathbf{x}_1,\mathbf{x}_2,\alpha)}\right. \nonumber \\  &+ \left.\frac{\left(\mu_{g_j}(\mathbf{x}_1,\mathbf{x}_2,\alpha) - g_j^{(2)} \right)^2}{\sigma^2_{g_j}(\mathbf{x}_1,\mathbf{x}_2,\alpha)} \right\}
\label{appmle}
\end{align}
where $\mu_{I_j}(\cdot)$, $\sigma^2_{I_j}(\cdot)$, $\mu_{g_j}(\cdot)$ and $\sigma^2_{g_j}(\cdot)$ are the mean and variance of the  intensity $I$ and the $g^{(2)}$, as defined in  Eq.\eqref{apprxNormal}.

Finding the MLE using Eq.~\eqref{appmle} is still non-trivial, as the log-likelihood involves highly non-linear functions of the parameters. Lagarias \textit{et al.} used the Nelder-Mead simplex algorithm (NMSA) to solve an equivalent equation to Eq.~\eqref{appmle} \cite{lagarias1998convergence}. In our case, however, the nonlinearity of the likelihood function results in  multiple local minima.  As a result, the optimization results given by NMSA are highly dependent on the selection of initial guesses, from which the algorithm typically converges to a local (non-global) minimum. To mitigate this problem, we first roughly estimate the $\{\mathbf{x}_1,\mathbf{x}_2,\alpha\}$ using a (1-st order) method of moments estimator (MME)~\cite{wooldridge2001applications}, i.e. 
\begin{align}
\ell(\hat{\mathbf{x}}_1^{0},\hat{\mathbf{x}}_2^{0},\hat{\alpha}^{0})=\arg\max \sum_{j=1}^{m}\left\{ \left( \mu_{I_j}(\mathbf{x}_1,\mathbf{x}_2,\alpha) - I_j\right)^2 \right.\nonumber \\
+ \left.\left(\mu_{g_j}(\mathbf{x}_1,\mathbf{x}_2,\alpha) - g_j^{(2)} \right)^2 \right\},
\label{MME}
\end{align}
as the seed for the NMSA. 

Since the objective function in the MME described in Eq.~\eqref{MME} is simpler than the likelihood function in terms of nonlinearity, the application of MME in this case  is less sensitive to the initial guess.
The final estimate is found  by applying NMSA to find the MLE  using the estimate given by the MME as the initial guesses.

\section{Effective PSF Characterisation}
\label{sec:weff}

In an optical system we have a point spread function (PSF), which is typically defined by the physical properties of the optical arrangement and therefore to determine parameters like the uncertainty in position. Although in our study there is no PSF in this definition, there are probabilities associated with it and that makes it possible to develop an \emph{effective PSF} that plays the same role on a statistical basis, i.e. where are the emitters most likely to be and how this probability varies as a function of space. However in any given simulation of our study, all we obtain is a single estimate of the emitter locations and their relative brightness. We use MC simulations to localise two emitters close with respect to the diffraction limit. After the MC simulation we use equal-cluster $k$-means  to group the inferred positions into two clusters and determine their centroids \cite{matlab2023equalkmeans}. These centroids of the two \textit{k}-means clusters are the estimated emitter locations. We define the \emph{effective PSF} size by determining the radial distance between each cluster centroid and the $[1-1/\sqrt{e}]^{\mathrm{th}}$ of MC inferred locations, $\rho_{39.5\%}$. If this distribution were exactly Gaussian, without any noise fluctuations, then $\rho_{39.5\%}$ would correspond to the
standard deviation of the PSF $w_0$, with $2\rho_{39.5\%}$ being the width of the PSF. We then determine the volume that includes the closest $\rho_{39.5\%}$ of the cluster centroids: 
\begin{align}
V_{39.5\%} = \frac{4}{3}\pi r^3
\label{eq:vol}
\end{align}

The \emph{effective width} of the point spread function $\bar{w}_{\mathrm{eff}}$ is the average of two emitter estimations $w_{\mathrm{eff},i}$:
\begin{align}
w_{\mathrm{eff},i} = 2r = \sqrt[3]{\frac{6V_{\sigma,i}}{\pi}},\\
\bar{w}_{\mathrm{eff}} = \frac{w_{\mathrm{eff,1}} + w_{\mathrm{eff,2}}}{2}
\label{eq:weff}
\end{align}

\begin{figure}[tb]
    \centering
    \includegraphics[width=\textwidth]{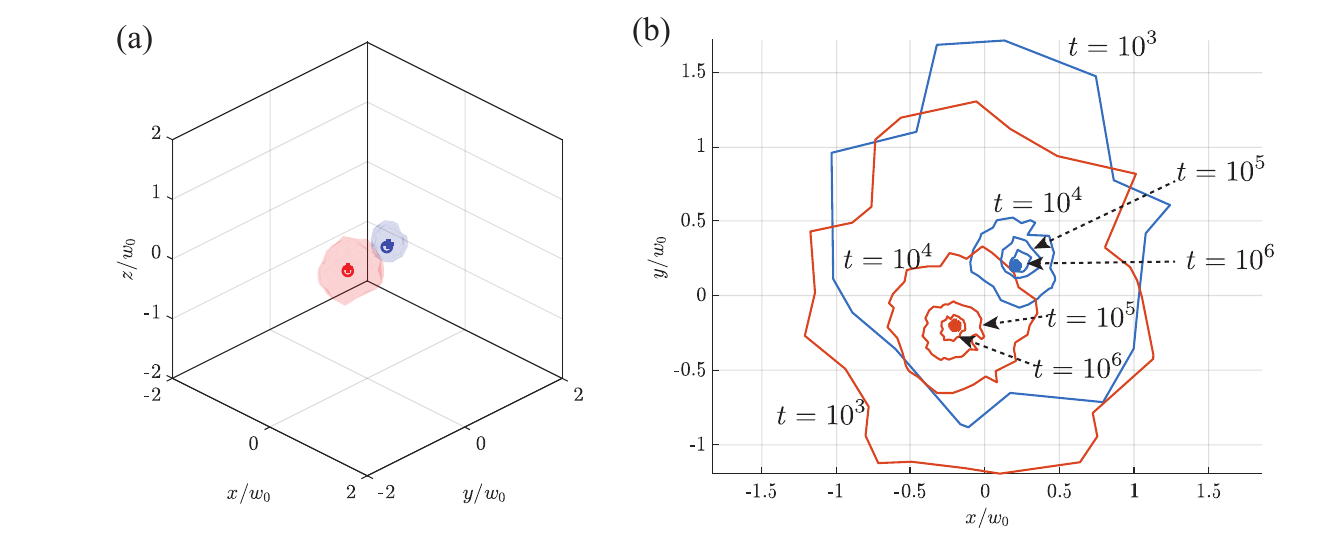}
    \caption{The 3D evolution of the inferred PSF.  The subfigure shows snapshot of the localisation at (a) $t = 10^4$ shown by the blue-shaded region for emitter 1 and red-shaded region for emitter 2. (b) The projection of inferred localisation on the $xoy$ plane, using a measurement configuration with four detector focal points arranged at the vertices of a tetrahedron, i.e. $\mathbf{\xi}_1 = (1, 1, 1)w_0$, $\mathbf{\xi}_2 = (-1, -1, 1)w_0$, $\mathbf{\xi}_3 = (-1, 1, -1)w_0$, and $\mathbf{\xi}_4 = (1, -1, -1)w_0$. The ground truth of the emitter positions are $\mathbf{x_1}=(0.2,0.2,0.2)w_0$ and $\mathbf{x_2}=(-0.2,-0.2,-0.2) w_0$, shown as blue and red stars inside the contours, with intrinsic brightness ratio $\alpha=0.5$. After 1000 MC simulations, contours show the uncertainties associated with the acquisition time $t=10^3, 10^4, 10^5,10^6$. The blue contours show the effective point-spread function for emitter 1, and the red contours show the effective point-spread function for emitter 2. $x$ and $y$ axes are the coordinates of the $xoy$ plane, normalized by the standard deviation of the PSF $w_0$.}
    \label{fig:contours}
\end{figure}

Fig.~\ref{fig:contours}(a) shows the volumes enclosed the $\rho_{39.5\%}$ of each emitter at acquisition $t=10^4$. To make the visualization clear we project the 3D image into the $xoy$ plane as contours, shown as Fig.\ref{fig:contours}(b). The contours are the projection of the closest $\rho_{39.5\%}$ inferred locations defined by Eq.~\eqref{eq:vol},  and the blue and red ones corresponding to emitter 1 and 2 respectively. It shows the evolution of  $\bar{w}_{\mathrm{eff}}$ with photon acquisition time $t$, where  $t$ is in  units of expected time between detections, assuming only one photon is fired from each emitter within one detection time. As $t$ increases more photons,  that is,  information,  would be measured. The figure shows that as detection time increases the one standard deviation contours converge to the ground truth locations at $\mathbf{x_1}=(0.2,0.2,0.2)w_0$ and $\mathbf{x_2}=(-0.2,-0.2,-0.2)w_0$.


\section{Localisation results of different detection configurations}
\label{sec:resluts}
\begin{figure}[tb]
    \centering
    \includegraphics[width=\textwidth]{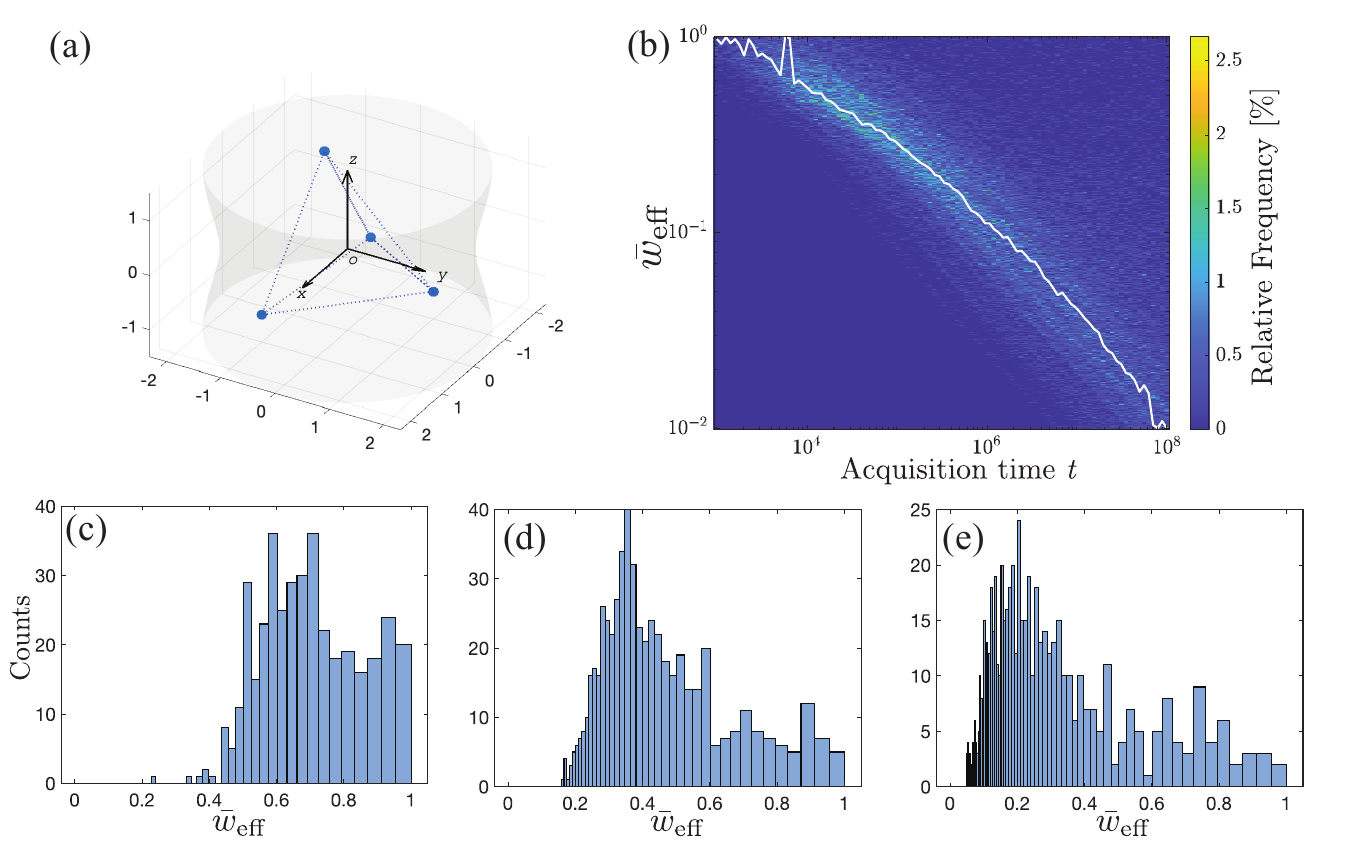}
    \caption{ Minimum four measurement locations are required to localize two emitters in 3D, (a) shows one tetrahedroal configuration design, with four focal points positioned at the vertices of a tetrahedron with $\mathbf{\xi}_1 = (1, 1, 1)w_0$, $\mathbf{\xi}_2 = (-1, -1, 1)w_0$, $\mathbf{\xi}_3 = (-1, 1, -1)w_0$, and $\xi_4 = (1, -1, -1)w_0$. The gray surface indicates the divergence of the Gaussian beam from the microscope objective. (b) Shows the $\bar{w}_{\mathrm{eff}}$ histogram map of 600 random ground truths and  1000 MC simulations per ground truth. (c)-(e) Show individual $\bar{w}_{\mathrm{eff}}$ histograms for times of $t = 10^4, 10^5, 10^6$, respectively.}
\label{fig:4det}
\end{figure}

To explore a large state space and to determine the effectiveness of our approach in generality, we now study the model performance with multiple random ground truths. To localise two emitters with seven variables $(x_1,y_1,z_1,x_2,y_2,z_2,\alpha)$, the minimum measurements required is four with both intensity and correlation information in each, therefore we start with the simplest detection in 3D with four detection focal points in a tetrahedral structure [Fig.\ref{fig:4det} (a)]. For each acquisition time $t$ we study $N=600$ random ground truths (randomly generated two emitter locations and relative intrinsic brightness), and run $M=1000$ MC simulations for each case/ground truth. In the map shown in Fig.~\ref{fig:4det}(b), each column is a histogram of $\bar{w}_{\mathrm{eff}}$ under a certain acquisition time, for example Fig.~\ref{fig:4det} (c)-(e) show the histogram binning in log scale at $t = 10^4, 10^5, 10^6$. We define relative frequency as the ratio of the bin height over the total number of experiments
\begin{equation}
    f_{relative} = \frac{n}{N\times M},
\end{equation}
where $n$ is the bin value. The histograms show the probability of the model giving the inferred locations with a specific $\bar{w}_{\mathrm{eff}}$. The mode of each histogram is shown as the white solid line overlaying on the histogram map in Fig.~\ref{fig:4det}(b). From the histograms we can see that as the acquisition time increases, the peak i.e. the mode of the histogram gradually shifts to smaller $\bar{w}_{\mathrm{eff}}$, indicating that the resolution of the model increases with the acquisition time $t$. 

\begin{figure}[tb]
    \centering
    \includegraphics[width=\textwidth]{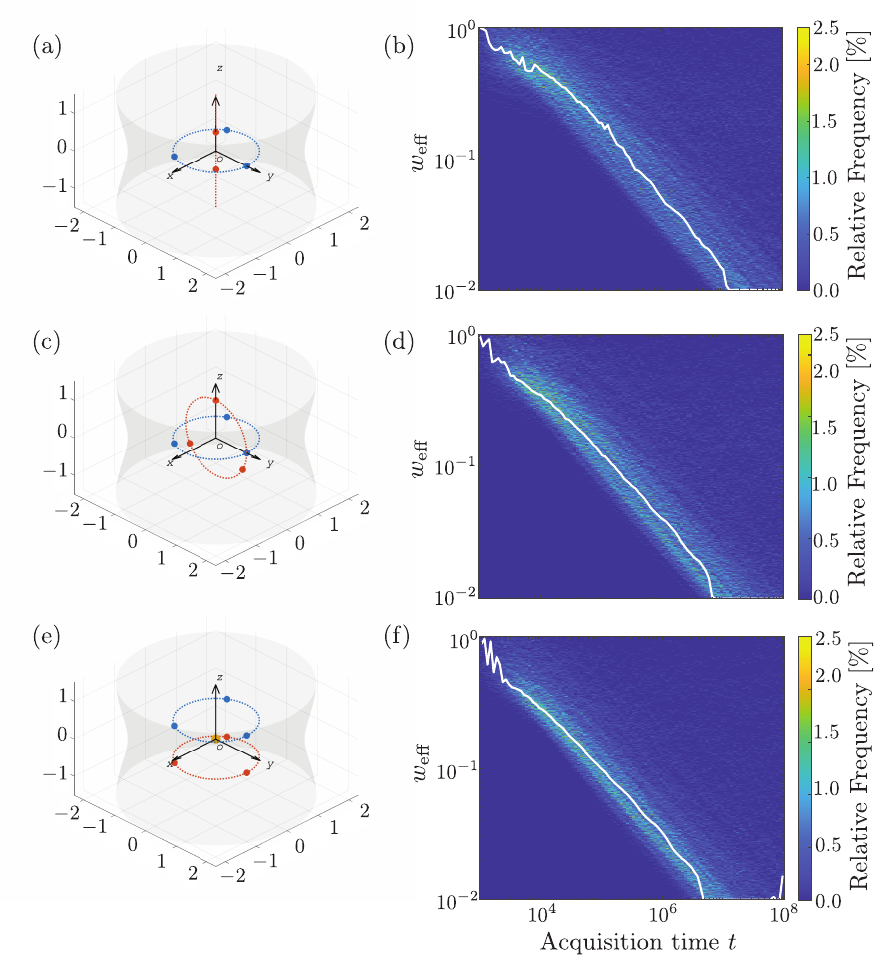}
    \caption{Three of the detection configurations analysed, with corresponding $w_{\text{eff}}$ histograms. (a) z-augmented trilateration configuration, with three focal points in the $xoy$ plane as a two-dimensional trilateration  and two along $z$ axis. (c) Orthogonal trilateration configuration with two perpendicular trilaterations. (e) Orthogonal trilateration plus: seven focal points with two parallel trilaterations plus one at the origin. (b) (d) and (f) are the corresponding $\bar{w}_{\mathrm{eff}}$ histogram maps of the right column configurations, with 600 random ground truths and 1000 MC simulations to localise each ground truth at each time point.}
    \label{fig:5-7det}
\end{figure}

\begin{figure}[tb]
    \centering
   \includegraphics[width=\textwidth]
   {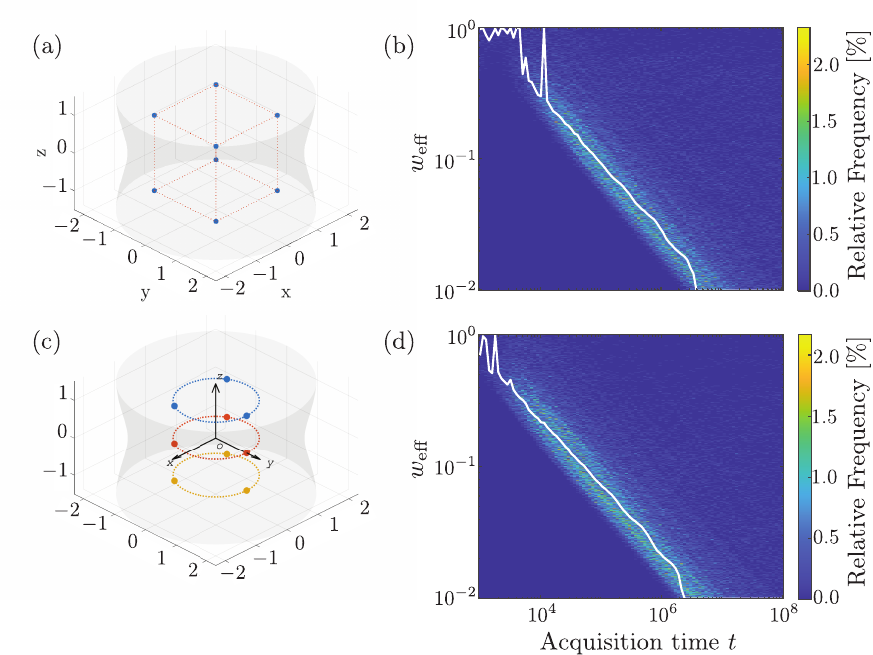}
    \caption{Two of the grid/z-scan type detection configurations with corresponding $w_{\text{eff}}$ histograms. (a) A grid $2 \times 2 \times 2$ configuration of eight focal points, with corresponding $w_{\text{eff}}$ histogram (b). (c) z-stack configuration: three scanned trilateration-type focal points to yield nine focal points with corresponding $w_{\text{eff}}$histogram. As expected, the increased number of measurement locations leads to increased resolution, although the ultimate scaling is the same. In each case the histogram maps were generated using  600 random ground truths and 1000 MC simulations per ground truth per time.}
    \label{fig:8-9det}
\end{figure}

There is considerable freedom to choose the focal points, $\mathbf{\xi_j}$, and each configuration will give rise to different resolution, with different results for different emitter system.  For this reason, we  explore several 3D measurement configurations. Motivated by the idea that the measurements could be realized in practice via two sets of two-dimensional HBT trilateration, we present two six-detector configurations [Fig.~\ref{fig:5-7det}(c)], or z-scan microscopy [Fig.~\ref{fig:8-9det}(c)]. Figs.~\ref{fig:5-7det} and ~\ref{fig:8-9det} show different measurement configurations identified by the positions of the maximum of the microscope confocal point spread function as it is translated through space.  These maxima are indicated by the blue dots in the schematic. From Figs.~\ref{fig:5-7det} and ~\ref{fig:8-9det} we gradually increase the number of detections. Right columns of Figs.~\ref{fig:5-7det} and ~\ref{fig:8-9det} show the histograms and fitting results for the mode of the distribution for the various measurement configurations in 3D space.

\section{Scaling law of resolution}
\label{sec:scaling}
\begin{figure}[tb]
    \centering
    \includegraphics[width=0.8\textwidth]{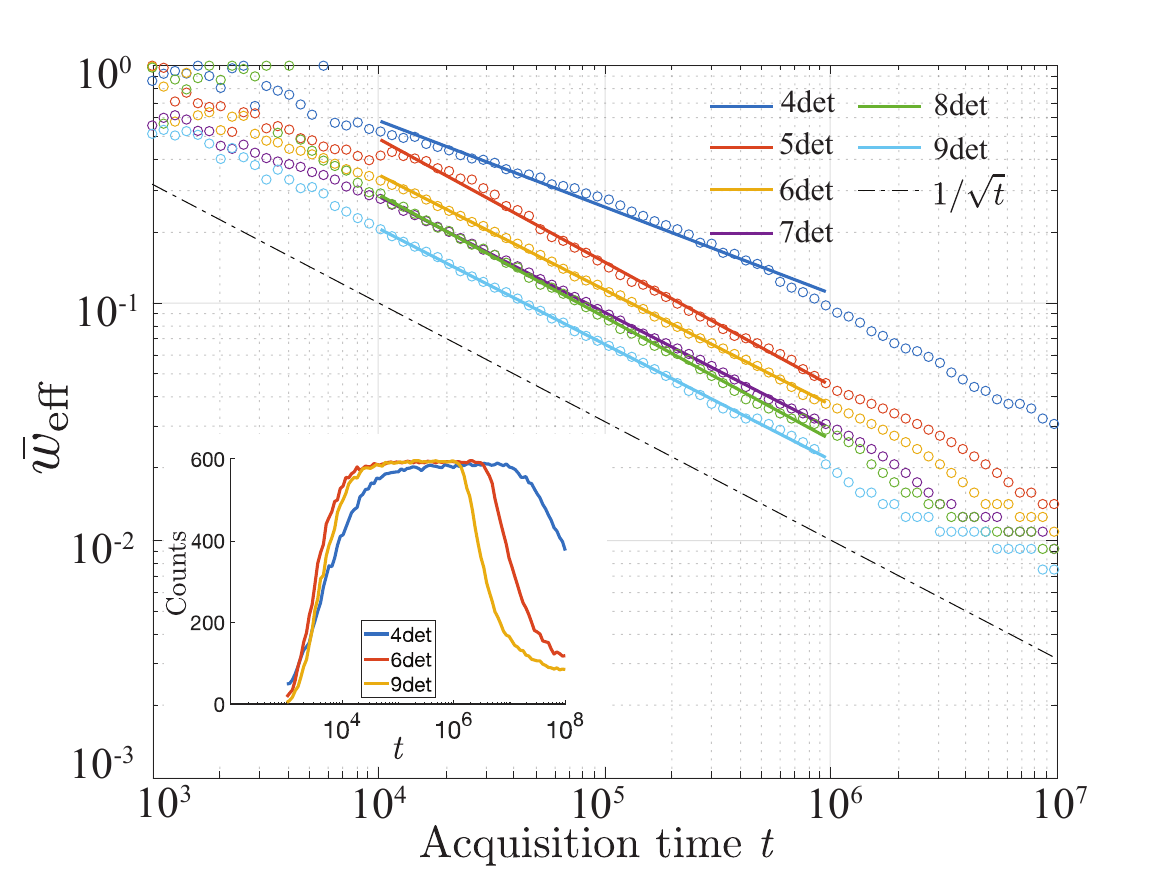}
     \caption{Scaling of the mode of the histogram of $\bar{w}_{\mathrm{eff}}$ as a function of total acquisition time. The dots are the modes of the histograms in Figs.\ref{fig:5-7det} and \ref{fig:8-9det}, and the lines are linear fitting. The inset shows the time window we choose to fit, within which the model could achieve sub-diffraction localisation for the majority of 600 cases.}
    \label{fig:scalingfit}
\end{figure}

To explore the resolution $\bar{w}_\mathrm{eff}$ scaling with the acquisition time $t$, we extract the modes of the histograms. The inset Fig.~\ref{fig:scalingfit} shows the total counts of the histogram as the function of $t$ (here we take the detection configurations with four, six, and nine detections as examples). We investigate the histogram with the $\bar{w}_\mathrm{eff}$ limitation $[0,1]w_0$, when a count falls into this region the model successfully achieves localisation resolution under the diffraction limit. The plateau shows that the majority $\bar{w}_{\mathrm{eff}}$ of the 600 random cases are localised with the sub-diffraction resolution. In the short time regime, the counts are low because the low correlation events are insufficient to realize sub-diffraction localisation. And when $t>10^6$, we suspect that limited by the computation power the MC statistics ($M=1000$ in this study) is not enough to realise the localisation. Therefore, we study the scaling law within the plateau region in Fig.~\ref{fig:scalingfit} inset, roughly $10^4$ to $10^6$.

The histogram could show a multimodal distribution because of the noise in the data and the histogram binning, therefore we fit the histogram and read out the peak as the mode. We then run the linear fit of the modes $\bar{w}_\mathrm{eff}$ and $t$ in log-log scale as Eq.\eqref{eq:loglogfit}, which gives the scaling law shown in Eq.\eqref{eq:scalinglaw}.
\begin{equation}
    \log \bar{w}_{\mathrm{eff}} = a \log t + b,
\label{eq:loglogfit} 
\end{equation}
\begin{equation}
    \bar{w}_{\mathrm{eff}} = 10^{a}*t^{b}
\label{eq:scalinglaw}
\end{equation}

Fig.~\ref{fig:scalingfit} shows the modes of histograms $\bar{w}_{\mathrm{eff}}$ as scatter dots and the fitting is shown as the solid lines. The scaling law fitting results are shown in Table \ref{table:scaling}. According to \cite{hemmer2012jo} the super-resolution enhancement goes with a $1/\sqrt{\mathrm{N_{photon}}}$ scaling, where $\mathrm{N_{photon}}$ is the total number of photons collected in the entire image spot during the observation time window. Because $\mathrm{N_{photon}}$ is proportional to the acquisition time, here we are expecting the resolution scaling law $1/\sqrt{t}$, so that the slope in Eq.~\eqref{eq:loglogfit} $b \approx -0.5$. Results in Fig.~\ref{fig:scalingfit} show that with fewer detections i.e. minimum four, the scaling is worse than $1/\sqrt{t}$, and increasing the detection numbers could improve the scaling therefore better localisation. The possible causes of the scaling better than $1/\sqrt{t}$ come from both the fitting of the histogram modes and the linear fitting, and the binning in histograms would also change the value of $b$ slightly, however it does not contradict that the theoretical precision scaling with the number of photons collected.

\begin{table}[tb]
\centering
\caption{The scaling law fitting results of different detection configurations. }
\begin{tabular}{ c| c|c|c|m{8cm} |c } 
\toprule[1pt]\midrule[0.3pt]
\makecell{Configuration} & \makecell{Number \\ focal \\ points}  & \makecell{Intercept \\ $a$} & \makecell{Slope \\ $b$} & Focal point coordinates, $\mathbf{\xi}_j/w_0$ & Figure \\
    \hline
    Tetrahedral & 4 & 1.2264 & -0.3644 & \makecell{$(1, 1, 1)$, $(-1, -1, 1)$,\\ $(-1, 1, -1)$, $(1, -1, -1)$} & \ref{fig:4det}(a)\\ 
    \hline
    \makecell{z-augmented \\ trilateration}  & 5  & 1.7728 & -0.5201 & \makecell{$(0,1,0)$, $(0,0,0.5)$, $(0,0,-0.5)$, \\ 
     $(-\sin(\pi/3),-\cos(\pi/3),0)$, \\ $(\sin(\pi/3),-\cos(\pi/3),0)$} & \ref{fig:5-7det}(a)\\ 
     \hline
    \makecell{Orthogonal \\ trilateration} & 6 & 1.4798 & -0.4849 & \makecell{$(0,1,0)$, $(0,0,1)$,  $(-\sin(\pi/3),-\cos(\pi/3),0)$, \\ $(\sin(\pi/3),-\cos(\pi/3),0)$, \\ $(0,-\sin(\pi/3),-\cos(\pi/3))$, \\ $(0,\sin(\pi/3),-\cos(\pi/3))$} & \ref{fig:5-7det}(c)\\ 
    \hline
    \makecell{Orthogonal \\ trilateration \\ plus} & 7 & 1.4063 & -0.4892 & \makecell{$(0,1,0.5)$, $(0,1,-0.5)$, $(0,0,0)$, \\ $(-\sin(\pi/3),-\cos(\pi/3),0.5)$,\\ $(\sin(\pi/3),-\cos(\pi/3),0.5)$, \\ $(-\sin(\pi/3),-\cos(\pi/3),-0.5)$, \\ $(\sin(\pi/3),-\cos(\pi/3),-0.5)$} & \ref{fig:5-7det}(e)\\ 
    \hline
    grid $2\times 2 \times 2$ &8 & 1.5222 & -0.5163 & \makecell{$(-1,-1,-1)$, $(-1,1,-1)$, $(1,-1,-1)$, \\ $(1,1,-1)$, $(-1,-1,1)$, $(-1,1,1)$,\\ $(1,-1,1)$, $(1,1,1)$} & \ref{fig:8-9det}(a) \\ 
    \hline
    z-stack &9 & 1.2658 & -0.4880 & \makecell{$(0,1,1)$, $(0,1,0)$,  $(0,1,-1)$,\\ $(-\sin(\pi/3),\cos(\pi/3),1)$,\\ $(\sin(\pi/3),-\cos(\pi/30),1)$,\\  $(-\sin(\pi/3),-\cos(\pi/3),0)$,\\ $(\sin(\pi/3),-\cos(\pi/30),0)$,\\ $(-\sin(\pi/3),-\cos(\pi/3),-1)$,\\ $(\sin(\pi/3),-\cos(\pi/3),-1)$} & \ref{fig:8-9det}(c) \\ 
\midrule[0.3pt]\bottomrule[1pt]
\end{tabular}
\label{table:scaling}
\end{table}


\section{Conclusion}
In summary, we have extended the QCM technique into three-dimensional localisation, by showing the localisation of two single photon emitters with unknown relative brightness.  Our results show  diffraction unlimited localisation with  increasing the detection time. We develop an approximated likelihood function as the analysis tool, and 
investigate several practical detection configurations by applying the estimator to many random ground truths, and the modes in statistics show the majority of the MC simulations generate $1/\sqrt(t)$ scaling in localisation precision. 

With the combination of intensity and second-order correlation information, our approach shows that it is achievable to localise sub-diffraction single-photon-emitters in three dimensions. We demonstrate the localisation of only two emitters, but the same approach could possibly be extended to more and ultimately track multiple emitters in three dimensions. The configurations that we have considered are relatively natural, and require only minor modification of a more conventional z-stack style confocal scan, albeit with relatively few focal points as the minimum number of focal points is four, in two focal planes. We also notice that the precision could be improved by increasing the measurements or adopting more efficient detection configurations, which could possibly be optimised via the investigation of Fisher information matrix in 3D.

\section*{Acknowledgement}
This work is funded by the Air Force Office of Scientific Research (FA9550-20-1-0276). ADG and BCG also acknowledges funding from the Australian Research Council (CE140100003
).

\bibliographystyle{unsrt}
\bibliography{references.bib}

\begin{thebibliography}{10}

\bibitem{betzig2006science}
Eric Betzig, George~H. Patterson, Rachid Sougrat, O.~Wolf Lindwasser, Scott
  Olenych, Juan~S. Bonifacino, Michael~W. Davidson, Jennifer
  Lippincott-Schwartz, and Harald~F. Hess.
\newblock Imaging intracellular fluorescent proteins at nanometer resolution.
\newblock {\em Science}, 313(5793):1642--1645, 2006.

\bibitem{hell2015review}
Stefan~W. Hell.
\newblock Nobel lecture: Nanoscopy with freely propagating light.
\newblock {\em Reviews of Modern Physics}, 87(4):1169--1181, 2015.
\newblock RMP.

\bibitem{koenderink2022nano}
A.~Femius Koenderink, Roman Tsukanov, Jörg Enderlein, Ignacio Izeddin, and
  Valentina Krachmalnicoff.
\newblock Super-resolution imaging: when biophysics meets nanophotonics.
\newblock {\em Nanophotonics}, 11(2):169--202, 2022.

\bibitem{heintzmann2017chemreview}
Rainer Heintzmann and Thomas Huser.
\newblock Super-resolution structured illumination microscopy.
\newblock {\em Chemical Reviews}, 117(23):13890--13908, 2017.

\bibitem{tenne2019np}
Ron Tenne, Uri Rossman, Batel Rephael, Yonatan Israel, Alexander
  Krupinski-Ptaszek, Radek Lapkiewicz, Yaron Silberberg, and Dan Oron.
\newblock Super-resolution enhancement by quantum image scanning microscopy.
\newblock {\em Nature Photonics}, 13(2):116--122, 2019.

\bibitem{monticone14prl}
D.~Gatto~Monticone, K.~Katamadze, P.~Traina, E.~Moreva, J.~Forneris,
  I.~Ruo-Berchera, P.~Olivero, I.~P Degiovanni, G.~Brida, and M.~Genovese.
\newblock Beating the {A}bbe diffraction limit in confocal microscopy via
  nonclassical photon statistics.
\newblock {\em Physical Review Letters}, 113(14):143602, 2014.

\bibitem{bartels2022ic}
Randy~A. Bartels, Gabe Murray, Jeff Field, and Jeff Squier.
\newblock Super-resolution imaging by computationally fusing quantum and
  classical optical information.
\newblock {\em Intelligent Computing}, 2022:0003, 2022.

\bibitem{nehme2018optica}
Elias Nehme, Lucien~E Weiss, Tomer Michaeli, and Yoav Shechtman.
\newblock Deep-{STORM}: super-resolution single-molecule microscopy by deep
  learning.
\newblock {\em Optica}, 5(4):458--464, 2018.

\bibitem{klauss2017springer}
André Klauss and Carsten Hille.
\newblock {\em Diffraction-Unlimited Fluorescence Imaging with an EasySTED
  Retrofitted Confocal Microscope}, pages 29--44.
\newblock Springer New York, New York, NY, 2017.

\bibitem{pascucci2019nc}
M.~Pascucci, S.~Ganesan, A.~Tripathi, O.~Katz, V.~Emiliani, and M.~Guillon.
\newblock Compressive three-dimensional super-resolution microscopy with
  speckle-saturated fluorescence excitation.
\newblock {\em Nature Communications}, 10(1):1327, 2019.

\bibitem{Aga1984}
D~A Agard.
\newblock Optical sectioning microscopy: Cellular architecture in three
  dimensions.
\newblock {\em Annual Review of Biophysics and Bioengineering}, 13(1):191--219,
  1984.
\newblock PMID: 6742801.

\bibitem{beaulieu2020natmethods}
Devin~R. Beaulieu, Ian~G. Davison, Kıvılcım Kılıç, Thomas~G. Bifano, and
  Jerome Mertz.
\newblock Simultaneous multiplane imaging with reverberation two-photon
  microscopy.
\newblock {\em Nature Methods}, 17(3):283--286, 2020.

\bibitem{abrahamsson2013natmethods}
Sara Abrahamsson, Jiji Chen, Bassam Hajj, Sjoerd Stallinga, Alexander~Y.
  Katsov, Jan Wisniewski, Gaku Mizuguchi, Pierre Soule, Florian Mueller,
  Claire~Dugast Darzacq, Xavier Darzacq, Carl Wu, Cornelia~I. Bargmann,
  David~A. Agard, Maxime Dahan, and Mats G.~L. Gustafsson.
\newblock Fast multicolor {3D} imaging using aberration-corrected multifocus
  microscopy.
\newblock {\em Nature Methods}, 10(1):60--63, 2013.

\bibitem{stelzer2015natmeth}
Ernst H.~K. Stelzer.
\newblock Light-sheet fluorescence microscopy for quantitative biology.
\newblock {\em Nature Methods}, 12(1):23--26, 2015.

\bibitem{swoger2014coldspring}
Jim Swoger, Francesco Pampaloni, and EH~Stelzer.
\newblock Light-sheet-based fluorescence microscopy for three-dimensional
  imaging of biological samples.
\newblock {\em Cold Spring Harb Protoc}, 2014(1):1--8, 2014.

\bibitem{stelzer2021natreview}
Ernst H.~K. Stelzer, Frederic Strobl, Bo-Jui Chang, Friedrich Preusser, Stephan
  Preibisch, Katie McDole, and Reto Fiolka.
\newblock Light sheet fluorescence microscopy.
\newblock {\em Nature Reviews Methods Primers}, 1(1):73, 2021.

\bibitem{szalai2021nc}
Alan~M. Szalai, Bruno Siarry, Jerónimo Lukin, David~J. Williamson, Nicolás
  Unsain, Alfredo Cáceres, Mauricio Pilo-Pais, Guillermo Acuna, Damián
  Refojo, Dylan~M. Owen, Sabrina Simoncelli, and Fernando~D. Stefani.
\newblock Three-dimensional total-internal reflection fluorescence nanoscopy
  with nanometric axial resolution by photometric localization of single
  molecules.
\newblock {\em Nature Communications}, 12(1):517, 2021.

\bibitem{axelrod2001traffic}
Daniel Axelrod.
\newblock Total internal reflection fluorescence microscopy in cell biology.
\newblock {\em Traffic}, 2(11):764--774, 2001.

\bibitem{boulanger2014pans}
Jérôme Boulanger, Charles Gueudry, Daniel Münch, Bertrand Cinquin, Perrine
  Paul-Gilloteaux, Sabine Bardin, Christophe Guérin, Fabrice Senger, Laurent
  Blanchoin, and Jean Salamero.
\newblock Fast high-resolution {3D} total internal reflection fluorescence
  microscopy by incidence angle scanning and azimuthal averaging.
\newblock {\em Proceedings of the National Academy of Sciences},
  111(48):17164--17169, 2014.

\bibitem{worboys20pra}
Josef~G. Worboys, Daniel~W. Drumm, and Andrew~D. Greentree.
\newblock Quantum multilateration: Subdiffraction emitter pair localization via
  three spatially separate {H}anbury {B}rown and {T}wiss measurements.
\newblock {\em Physical Review A}, 101(1):013810, 2020.

\bibitem{HSH1995}
Stefan~W Hell, Jori Soukka, and Pekka~E Hänninen.
\newblock Two- and multiphoton detection as an imaging mode and means of
  increasing the resolution in far-field light microscopy: A study based on
  photon-optics.
\newblock {\em Bioimaging}, 3(2):64--69, 1995.

\bibitem{israel2017nc}
Yonatan Israel, Ron Tenne, Dan Oron, and Yaron Silberberg.
\newblock Quantum correlation enhanced super-resolution localization microscopy
  enabled by a fibre bundle camera.
\newblock {\em Nature Communications}, 8:14786, 2017.

\bibitem{schwartz2013nl}
Osip Schwartz, Jonathan~M. Levitt, Ron Tenne, Stella Itzhakov, Zvicka Deutsch,
  and Dan Oron.
\newblock Superresolution microscopy with quantum emitters.
\newblock {\em Nano Letters}, 13(12):5832--5836, 2013.

\bibitem{schwartz2012pra}
O.~Schwartz and D.~Oron.
\newblock Improved resolution in fluorescence microscopy using quantum
  correlations.
\newblock {\em Physical Review A}, 85(3):033812, 2012.
\newblock PRA.

\bibitem{simon2010oe}
D.~S. Simon and A.~V. Sergienko.
\newblock The correlation confocal microscope.
\newblock {\em Optics Express}, 18(10):9765--9779, 2010.

\bibitem{IWW+2017}
Jaroslav Icha, Michael Weber, Jennifer~C. Waters, and Caren Norden.
\newblock Phototoxicity in live fluorescence microscopy, and how to avoid it.
\newblock {\em BioEssays}, 39(8):1700003, 2017.

\bibitem{SCL+2014}
Romana Schirhagl, Kevin Chang, Michael Loretz, and Christian~L. Degen.
\newblock Nitrogen-vacancy centers in diamond: Nanoscale sensors for physics
  and biology.
\newblock {\em Annual Review of Physical Chemistry}, 65(1):83--105, 2014.
\newblock PMID: 24274702.

\bibitem{lagarias1998convergence}
Jeffrey~C Lagarias, James~A Reeds, Margaret~H Wright, and Paul~E Wright.
\newblock Convergence properties of the {N}elder--{M}ead simplex method in low
  dimensions.
\newblock {\em SIAM Journal on optimization}, 9(1):112--147, 1998.

\bibitem{wooldridge2001applications}
Jeffrey~M Wooldridge.
\newblock Applications of generalized method of moments estimation.
\newblock {\em Journal of Economic perspectives}, 15(4):87--100, 2001.

\bibitem{matlab2023equalkmeans}
Kai-Daniel Büchter.
\newblock Algorithm to generate equal-sized clusters based on k-means, 2023.
\newblock MATLAB Central File Exchange. Retrieved August 25, 2023.

\bibitem{hemmer2012jo}
Philip~R. Hemmer and Todd Zapata.
\newblock The universal scaling laws that determine the achievable resolution
  in different schemes for super-resolution imaging.
\newblock {\em Journal of Optics}, 14(8):083002, 2012.

\end{thebibliography}

\end{document}